\documentclass[a4paper,12pt,oneside,reqno]{amsart}
\usepackage{hyperref}
\usepackage[headinclude,DIV13]{typearea}
\areaset{15.1cm}{25.0cm}
\usepackage{txfonts,amssymb,amsmath,amsthm,bbm}

\usepackage[latin1] {inputenc}
\usepackage{graphicx, psfrag}
\usepackage{xcolor}
\usepackage{subfigure}

\newtheorem{theorem}{Theorem}[section]

\newtheorem{corollary}[theorem]{Corollary}
\newtheorem{assumption}[theorem]{Assumption}
\newtheorem{definition}[theorem]{Definition}

\definecolor{bbm}{RGB}{51,153,0}
\definecolor{above}{RGB}{128,0,128}
\definecolor{below}{RGB}{102,0,204}
\definecolor{cascade}{RGB}{204,0,0}
\definecolor{iid}{RGB}{153,51,0}

\theoremstyle{remark}
\newtheorem*{remark}{Remark}

\def\paragraph#1{\noindent \textbf{#1}}

\numberwithin{equation}{section}

\def\<{\langle}
\def\>{\rangle}

\def\f{\phi}

\def\L{\Lambda}

\let\cal=\mathcal

\def\NN{{\cal N}}

\def\PP{{\cal P}}

\def\RR{{\cal R}}

 \def \L {{\Lambda}}

 \def \f {{\phi}}

 \def \ba {\begin{array}}
 \def \ea {\end{array}}

 \def \cL {{\cal L}}

 \def \cX {{\cal X}}
 \newcommand{\be}{\begin{equation}}
 \newcommand{\ee}{\end{equation}}

\newcommand{\bea}{\begin{eqnarray}}
 \newcommand{\eea}{\end{eqnarray}}
\def\TH(#1){\label{#1}}\def\thv(#1){\ref{#1}}
\def\Eq(#1){\label{#1}}\def\eqv(#1){(\ref{#1})}

 \def \1{\mathbbm{1}}

\begin{document}

 \title[Stochastic models for adaptive dynamics]{Stochastic models for adaptive dynamics: Scaling limits and
diversity}
\author[A. Bovier]{Anton Bovier}
 \address{A. Bovier\\Institut f\"ur Angewandte Mathematik\\
Rheinische Friedrich-Wilhelms-Universität\\ Endenicher Allee 60\\ 53115 Bonn, Germany }
\email{bovier@uni-bonn.de}

\date{\today}

 \begin{abstract} 
 I discuss the so-called  stochastic individual based model of adaptive dynamics and in particular how different scaling limits can be obtained by taking limits of large populations, small mutation rate, and 
small effect of single mutations together with appropriate time rescaling. 
In particular, one derives the trait substitution sequence, polymorphic evolution sequence, and the canonical equation of adaptive dynamics. In addition, I show how the escape from an evolutionary stable conditions can occur as a metastable transition. This is a review paper that will appear in 
 "Probabilistic Structures in Evolution", ed. by E. Baake and A. Wakolbinger.
   \end{abstract}

\thanks{
This work was partly funded by the Deutsche Forschungsgemeinschaft (DFG, German Research Foundation) under Germany's Excellence Strategy - GZ 2047/1, Projekt-ID 390685813 and GZ 2151 - Project-ID 390873048,
through the Collaborative Research Center 1060 \emph{The Mathematics 
of Emergent Effects}
and through the Priority Programme 1590 \emph{Probabilistic Structures in Evolution}.
 }

\maketitle

\section{Theories of evolution}

The key features inherent in any biological system that are  driving 
forces of \emph{evolution}
 identified already by
Darwin  \cite{AB-Darwin1859}  are:
\begin{itemize}
\item \emph{birth and death}: individuals die and reproduce
\item \emph{heredity}: the offspring of individuals inherit properties 
(\emph{traits}) of their ancestors
\item \emph{mutation}: heredity is not perfect, and sometimes the traits of 
the offspring differ from those of their
ancestors
\item \emph{selection}, or \emph{the survival of the fittest}, a concept inspired by  the essay \emph{An 
Essay on the Principle of Population} by Thomas Malthus \cite{AB-malthus1798} from 1798. 
\end{itemize}

Selection results from the interaction between individuals, notably the competition for resources,
but also many other effects (predation, symbiosis, parasitism, etc.).

 The classical theories of 
 \emph{Evolution} can be broadly 
classified into two branches: \emph{population dynamics}, that focuses on \emph{ecology}, i.e. on aspects 
of competition and other interactions between different species, and \emph{population genetics},
that focuses on heredity and the genealogical structure of populations.

\subsection{Population dynamics}
Population dynamics can indeed be traced back to Malthus' essay \cite{AB-malthus1798}, where 
he lays out that an unrestrained population will grow exponentially, but that in all real \emph{states}
this growth must be restrained by the limited amounts of food that is available.  In modern terms, this 
leads to simplest differential equation describing the time evolution of  the size $n(t)$ of a (monomorphic) 
population
\begin{equation}\label{AB-lv.1}
\frac d{dt}n(t) = n(t) r - cn(t)^2,
\end{equation}
where $r=b-d$ is the difference between the birth-rate $b$ and the death-rate $d$, and $c$ is a measure 
of the competitive pressure two individuals exert on each other.
Note that here $n$ is not the number of individuals (which would need to be an integer), but 
rather a rescaled for the mass of the population when the mass of an individual tends to zero 
while the number of individuals tends to infinity at the same rate. 

The analysis of differential equations of this type goes back to the works of Alfred Lotka 
\cite{AB-lotka1912}
and Vito Volterra \cite{AB-volterra1928}.  One calls systems of differential equations of the form 
\begin{equation}\label{AB-lv.2}
\frac d{dt}n_i(t) = n_i(t) \left(r_i - \sum_{k=1}^dc_{ik}n_k(t)\right),  \quad i=1,\dots,d,
\end{equation}
\emph{competitive Lotka-Volterra equations} if all coefficients $c_{ik}$ are non-negative, and
simply Lotka-Volterra equations in the general case.

\subsection{Adaptive dynamics}
\emph{Adaptive dynamics} (AD) is somewhat of an outgrowth of both population dynamics and population 
genetics.   Hans Metz, one of its prominent protagonists 
(see \cite{AB-MG95} for one of the fundamental papers), describes it in his essay \emph{Adaptive 
Dynamics} \cite{AB-metz2012}
as ``a simplified theoretical approach to \emph{meso-evolution}, defined here as evolutionary changes in 
the values of traits of representative individuals and concomitant patterns of taxonomic diversification".
Further, ``Trait changes result
from the micro-evolutionary process of mutant substitutions taking place against the
backdrop of a genetic architecture and developmental system as deliverers of mutational
variation."
An important assumption of AD, that we will encounter in the analysis of the mathematical models later 
on, is the separation of the time scales of ecology and evolution. 
Note that this assumption implies an effectively low rate of (trait-changing, advantageous) mutations.

Adaptive dynamics thus deals with our fundamental objective, describing how populations (rather than 
individuals), characterised by some homogeneous traits, evolve in time into multifaceted families of 
populations exhibiting a broad variety of traits. A fundamental concept characterising 
this \emph{trait space} is \emph{fitness}. Fitness  is a complicates concept and  has different meanings. For an extensive discussion of this notion, see e.g. \cite{AB-MN92} or Chapter 8 of \cite{AB-ewens04}. In population dynamics, fitness is the initial exponential growth rate of a population; in population 
genetics, it is often referred to as the probability of an individual to reach maturity, i.e. to produce 
offspring. In adaptive dynamics, fitness is viewed in a more \emph{dynamic} fashion \cite{AB-MN92}. Fitness of a 
population with a specific trait does not only depend on this trait, but also on the state of the 
entire population, as effects of competition or other interactions play a significant rôle. 
In fact, if a population is in ecological equilibrium, then all co-existing traits have zero fitness (i.e. they 
do not grow or shrink). Under the assumption of separation of time scales, the 
only relevant fitness parameter is then the so-called \emph{invasion fitness}, 
which is the exponential growth rate of a mutant born with a given trait in the presence of 
the current equilibrium population.
 
 A resulting fundamental concept of adaptive dynamics is  that of an 
 \emph{evolutionary stable
 condition} (ESC). This is a population in  ecological equilibrium such that 
 all traits that are accessible by a single mutation from the current population  have negative 
 invasion fitness. The fate of evolution is  reached if a population is in 
 such a condition. 
 Clearly, for a given environment, there may be many ESCs (just think of two islands separated by an ocean that
 the cannot be crossed. Then, an ecological equilibrium on any one of the islands (with the other unpopulated)
 or on both islands is an ESC).
 The task of 
 adaptive dynamics would 
 then be to identify all ESCs and to decide the ways to reach such ESCs. 
 On the way towards 
 ESCs, adaptive dynamics identifies two kew mechanisms: the 
 \emph{canonical equation of 
 adaptive dyamics} (CEAD), which describes how a (monomorphic) 
 population moves in trait space 
 (under the further assumption of small mutation steps), and
 \emph{adaptive speciation}, which describes how polymorphism emerges 
 as a population arrives
 at a state where several directions of mutations are viable and a 
 bifurcation can happen.
 
 A particularly simple limit of adaptive dynamics is the regime of strong selection and weak mutation, which gives rise to
 so-called \textit{adaptive walks} \cite{AB-KaLe87,AB-Kau92,AB-Orr03}. Here, evolution is modelled as a random walk on the trait space that moves towards higher fitness as the population adapts to its environment. More precisely, a discrete state space is equipped with a graph structure that marks the possibility of mutation between neighbours. A fixed, but possibly random, fitness landscape is imposed on the trait space. In contrast to the above, this \textit{individual fitness} is not dependent on the current state of the population. Adaptive walks move along neighbours of increasing fitness, according to some transition law, towards a local or global optimum. For a survey on adaptive walks and the analysis of different fitness landscapes, see 
 the chapter by Joachim Krug 
 \cite{AB-Krug2019} in this volume.

\section[The individual based model]{The individual based model of adaptive dynamics}
In this section we describe the  basic stochastic, individual based model of adaptive dynamics. 
We consider a population that is composed of a number of individuals, each of them characterised by a  phenotypic trait that takes values in some (Polish) space $\mathcal X$.
This model was studied extensively, see, e.g. \cite{AB-C_ME,AB-Cha06,AB-CM11,AB-FM04}.  See also 
\cite{AB-BanMel2015} for a review.
The \emph{trait space} $\mathcal X$ can in principle be chooses quite arbitrarily, e.g. the model can  defined for
$\mathcal X$ being just a Polish space. In various circumstances, one may wish to put more structure on it. 
For the applications we discuss in these notes, it will either be a finite set or a  (closed) 
subset of $\mathbb {R}^d$, and for convenience we take $\mathcal X$ to be always a subset of $\mathbb R^d$
in the sequel.
The dynamics  is driven by the following key parameters, that are functions of the traits: 
\begin{itemize}
\item[(i)] {$b(x)\in\mathbb R_+$ is the \emph{reproduction rate} of an individual with trait $x\in\mathcal X$.}
\item[(ii)]{$d(x)\in\mathbb R_+$ is the \emph{rate of natural death} of an individual with trait $x\in\mathcal X$.}
\item[(iii)]{$c(x,y)\in\mathbb R_+$ is the \emph{competition kernel}
		which models the competition pressure felt by an individual with trait $x\in\mathcal X$ from an individual with trait $y\in\mathcal X$.}
\item[(iv)]{ $ m(x)\in [0,1]$ is the \emph{probability that a mutation occurs at birth} from an individual with trait $x\in\mathcal X$.}
    \item[(v)]{$M(x,dy)$ is the \emph{mutation law}. If the mutant is born from an individual with trait $x$, then the mutant trait  is given by $x+y\in \mathcal X$, where $y$ is a random variable with law $M(x,dy)$.  }
\end{itemize}

At any time $t$  we consider a finite number, $N_t$, of individuals with   trait value $x_i(t)\in \mathcal X$. 
The population state at time $t$  is represented by  point measures,
\begin{equation}
 	\nu _t=  \sum_{i=1}^{N_t}\delta_{x_i(t)}.
 \end{equation} 
Let $\langle \nu , f \rangle$ denote the integral of a measurable function $f$ with respect to the measure $\nu $. 
Then $\langle\nu _t,\1\rangle=N_t $ and for any $x\in\mathcal X$, 
the non-negative number $\langle\nu _t,\1_{\{x\}}\rangle$ is called the \emph{density of trait $x$ at time $t$}. 
Let $\mathcal M(\mathcal X)$ denote the set of finite nonnegative point measures on $\mathcal X$, equipped with the vague topology, 
\begin{equation}
	\mathcal M(\mathcal X)\equiv \left\{ \sum_{i=1}^{n}\delta_{x_i}\,:\, n\geq 0,\; x_1,...,x_n\in \mathcal X\right\}.
 \end{equation}
 The population process, $(\nu_t)_{t\geq0}$, is then defined as a 
$\mathcal M(\mathcal X)$-valued Markov process with  generator $\cL$, 
defined, for any bounded measurable function $f$ from $\mathcal M(\mathcal X)$ to $\mathbb R$ and for all $\nu\in \mathcal M(\mathcal X)$, by
\begin{eqnarray}\label{AB-generator.1}
 	\nonumber
(\cL f)(\nu) & = & \int_{\mathcal X}\biggl(f\Bigl(\nu + {\delta_x}\Bigr)-f(\nu )\biggr)\bigl(1- m(x)\bigr)b(x)\:\nu (dx)\\ \nonumber
						&&+\int_{\mathcal X}\int_{{\mathcal X}}\biggl(f\Bigl(\nu+{\delta_{x+y}} \Bigr)-f(\nu )\biggr) 			
							m(x)b(x)\:M(x,dy)\: \nu (dx)\\
						&&+\int_{\mathcal X}\biggl(f\Bigl(\nu -{\delta_x} \Bigr)-f(\nu )\biggr)\Bigl(d(x)+
						\int_{\mathcal X}c(x,y)\nu (dy)\Bigr)\:\nu (dx).
\end{eqnarray}
The first and second terms are linear (in $\nu$) and describe the births (without and with mutation), 
but the third term is non-linear and describes the deaths due to age or competition. 
The density-dependent non-linearity of the third term models the competition in the population, 
and hence drives the selection process.
\begin{assumption}\label{AB-ass}We make the following assumptions on the parameters of the model:
\begin{enumerate}
\renewcommand{\labelenumi}{(\roman{enumi})}
\item{$b$, $d$ and $c$ are measurable functions, and there exist $\overline b,\overline d,\overline c<\infty$ such that}
\begin{center}$0\leq b(.)\leq \overline b,\quad 0\leq d(.)\leq \overline d\quad$ and $\quad 0\leq c(.\:,.)\leq \overline c.$\end{center}
\item{There exists $\underline c>0$ such that for all $x\in\mathcal X$, 
$\underline c\leq c(x,x)$.}
\item{The support of $M(x,\:.\:)$ is uniformly bounded for all $x\in \mathcal X$. }
%
\end{enumerate} 
\end{assumption}
\begin{remark}
Assumptions (i)  allows to deduce the existence and uniqueness in law of a process on $\mathbb D(\mathbb R_+,\mathcal M(\mathcal X))$ with infinitesimal generator $ \cL$ (cf. \cite{AB-FM04}). 
Assumption (ii) ensures the population size to stay bounded locally. Assumption (iii)  is made in view of the 
convergence to the canonical equation, see below, and can be relaxed. 
\end{remark}

\section[Scaling limits]{Scaling limits}

There are three natural parameters that can be introduced into the model that give rise to interesting and biologically relevant scaling limits. 
These are 
\begin{itemize}
\item [(i)] The population size, or \emph{carrying capacity}, $K$. This is achieved by dividing the competition kernel $c$ by $K$, so that it requires of order $K$ individuals  to affect the death rate of one individual in a significant way. To obtain a limit then also requires to divide the measures $\nu$ by $K$. 
\item[(ii)] The mutation rate, $u$. Multiplying the mutation rate $m(x)$  by $u$ allows to study limits of small
mutation rates. 
\item[(iii)]  The effect of a single mutation step can be scaled to zero. The mutation step size can be scaled to zero by introducing a parameter $\sigma$ and 
replacing 
$\delta_{x+y} $ in the mutation term of the generator by $\delta_{x+\sigma y}$. 
\end{itemize}
The generator with these scaling parameters acting on the space of rescaled measures then reads
\begin{eqnarray}\label{AB-generator.1}
 	\nonumber
(\cL^K f)(\nu^K) &=& \int_{\mathcal X}\biggl(f\Bigl(\nu^K +{\textstyle 1\over K} {\delta_x}\Bigr)-f(\nu^K )\biggr)\bigl(1- um(x)\bigr)b(x)\:K\nu^K (dx)\\ \nonumber
						&&\hspace{-8mm}+\int_{\mathcal X}\int_{{\mathcal X}}\biggl(f\Bigl(\nu^K+{\textstyle 1\over K}{\delta_{x+\sigma y}} \Bigr)-f(\nu^K)\biggr) 			
							um(x)b(x)\:M(x,dy)\: K\nu^K (dx)\\
						&&\hspace{-8mm}+\int_{\mathcal X}\biggl(f\Bigl(\nu^K-
					{\textstyle 1\over K} {\delta_x} \Bigr)-f(\nu^K)\biggr)\Bigl(d(x)+
						\int_{\mathcal X}c(x,y)\nu^K (dy)\Bigr)\:K\nu^K (dx).
\end{eqnarray}
In general one is interested in taking limits as $K\uparrow\infty$, $u\downarrow 0$, and $\sigma\downarrow 0$.
At the same time, we may want to scale time in such a way to obtain interesting effects. Having large $K$, small 
$u$, and small $\sigma$ is biologically reasonable in many (but not all) situations. 

\subsection[Law of large numbers]{The law of large numbers}\label{AB-known_results}
A fundamental result, that also provides a frequently used tool, is a \emph{Law of Large Numbers (LLN)}, that asserts convergence of the process to a deterministic limit over finite time intervals when $K$ tends to infinity.   This LLN goes in fact back to Ethier and Kurtz \cite{AB-EthKur1986} in the case of finite trait space and was generalised by Fournier and  M\'el\'eard \cite{AB-FM04}. See also \cite{AB-BanMel2015}.

\begin{theorem}\label{AB-lil.1}
Fix $u$ and $\sigma$.
Let Assumption \ref{AB-ass} hold and assume in addition that 
the initial conditions $\nu_0^K$ converge,  as $K\uparrow\infty$, in law and for the weak topology on 
$\mathcal M(\mathcal X)$, to some deterministic finite measure $\xi_0\in \mathcal M(\mathcal X)$ 
and that $\sup_K \mathbb E\left [ \langle \nu^{K}_0, \1\rangle^3\right]<\infty$. 
%
Then, for all $T > 0$, the sequence $\nu^K$ converges, as $K\uparrow\infty$, in law, 
on the Skorokhod space $\mathbb D([0,T ],\mathcal M ( \mathcal X))$, to a deterministic continuous function 
$\xi\!\in\! C([0,T ],\mathcal M ( \mathcal X))$.
This measure-valued function $\xi$ is the unique solution, 
satisfying $\sup_{t\in[0,T ]}\langle \xi_t, \1\rangle \!<\infty $, 
of the integro-differential equation written in its weak form:
for all bounded and measurable functions, $h:\mathcal X \to\mathbb R$,
\begin{eqnarray}\label{lln.1}
&&\int_{\mathcal X}h(x) \xi_t(dx)
	-\int_{\mathcal X} h(x) \xi_0(dx) \\\nonumber
		&&=\!\int_0^t ds \int_{\mathcal X}u m(x)b(x) \int_{\mathbb Z} M(x,dy) h(x+\sigma y)\xi_s(dx) 
						  \\ \nonumber
	&&+\!\int_0^t  ds \int_{\mathcal X} h(x)\Big(\left(1\!-\!u m(x)\right)b(x)
							\!-\!d(x)\!-\!\int_{\mathcal X}\xi_s(dy)c(x,y) \Big)\xi_s(dx) .
\end{eqnarray}
\end{theorem}

\begin{remark} In all the results mentioned in these notes, the LLN is in fact only used for finite trait spaces. 
\end{remark}

\subsection{Scaling $u\downarrow 0$ in the deterministic limit}

In the absence of mutations ($u=0$) and if initially there exists a finite number of phenotypes in the population, 
one obtains   convergence to the competitive system of Lotka-Volterra equations defined below
 (see \cite{AB-FM04}).
\begin{corollary}[The special case $u=0$ and $\xi_0$ is $n$-morphic]\label{AB-cor}
If the same assumptions as in the theorem above with $u=0$ hold and if in addition $\xi_0=\sum_{i=1}^{n} z_i(0)\delta_{x_i}$, then $\xi_t$ is given by  
$\xi_t=\sum_{i=1}^{n}z_i(t)\delta_{x_i}$, where $z_i$ is the solution of the competitive system of Lotka-Volterra equations defined below.
\end{corollary}
\begin{definition}For any $(x_1,...,x_n)\in\mathcal X^n$, 
we denote by $LV(n,(x_1,...,x_n))$ the \emph{competitive system of Lotka-Volterra equations}  defined by
\begin{equation}\label{AB-LV-System}
\frac {d\:z_i(t)}{dt}= z_i(t)\biggl(b(x_i)-d(x_i)-\sum_{j=1}^n c(x_i,x_j)z_j(t)
\biggr), \qquad 1\leq i\leq n.
\end{equation}
\end{definition}
It is also  easy to see (using Gronwall's lemma) that on finite time intervals, solutions converge, as $u\downarrow 0$, to those of the system with $u=0$. The same is \emph{not} true if time tends to 
infinity as $u\downarrow 0$.

 We introduce the notation of coexisting traits and of invasion fitness (see \cite{AB-Cha06}).
\begin{definition}
The distinct traits $x$ and $y$ \emph{coexist} if the system $LV(2,(x,y))$ admits a unique non-trivial equilibrium, $\overline { z} (x,y)\! \in\! (0,\infty)^2$, which is locally strictly stable in the sense that the eigenvalues of the Jacobian matrix of the system $LV(2, (x,y))$ at $ \overline { z }(x,y)$ are all strictly negative. 
\end{definition}
The invasion of a single mutant trait in a monomorphic population which is close to its equilibrium is governed by its initial growth rate. Therefore, it is convenient to define the fitness of a mutant trait by its initial growth rate.
\begin{definition} If the resident population has the trait  $x\in\mathcal X$, then we call the following function \emph{invasion fitness} of the mutant trait $y$
\begin{align}
f(y, x)=b(y)-d(y)-c(y,x)\overline z(x).
\end{align}
\end{definition}
\begin{remark}
The unique strictly stable equilibrium of $LV(1,x)$ is $ \overline z(x)=\frac{b(x)-d(x)}{c(x,x)}$, and hence $f(x,x)=0$ for all $x\in\mathcal{X}$.
\end{remark} 

Determining polymorphic fixed points with a subset of $\ell \leq n$ components 
(take w.l.o.g. the first $\ell $ components, leads to the \emph{linear} equations 
 \begin{equation}\label{AB-equi.5}
b(x_i)-d(x_i)=\sum_{j=1}^\ell c(x_i,x_j)\bar z_j, \qquad 1\leq i\leq \ell,
\end{equation}
for the equilibrium values $\bar z_j$, which in addition must be all 
strictly positive. The existence of such equilibria clearly requires conditions 
on the parameters that are more difficult to verify.

\subsection{Small mutation limit at divergent time scales}
\label{AB-logscale}
We have seen that for finite time horizons, the limit of the deterministic equations as $u\downarrow 0$
is a mutation free ecological equation. The reason for this is that growth of solutions is at most 
exponential in time, and so anything seeded by a mutation term is proportional to $u$ and  will vanish in 
the limit. This is no longer true if we consider time scales that depend on $u$. To understand this, consider an initial 
condition that is monomorphic and the simplest case where  ${\mathcal X}$ is just the set $\{1,2\}$. Then the deterministic system can be reduced to the two-dimensional Lotka-Volterra system with mutation,  
(to lighten the notation we set $m(1)=m(2)=1$)
\begin{eqnarray}\label{AB-simex.1}
\frac {dz_1(t)}{dt}&=& z_1(t)(r_1-c(1,1)z_1(t)-c(1,2)z_2(t))-uM(1,2)z_1(t)+uM(2,1)z_2(t),\nonumber
\\
\frac {dz_2(t)}{dt}&=& z_2(t)(r_2-c(2,2)z_2(t)-c(2,1)z_1(t))-uM(2,1)z_2(t)+uM(1,2)z_1(t).\nonumber\\
\end{eqnarray}
Assume that $z_1(0)=\bar z_1\equiv \frac {r_1}{c(1,1)}$ and $z_2(0)=0$. 
Assume further that the invasion fitness of type two is positive, i.e. $r_2-c(2,1)\bar z_1>0$.
Then, at time $0$, we have 
\begin{eqnarray}\label{AB-simex.2}
\frac {dz_1(0)}{dt}&=& \bar z_1(r_1-c(1,1)\bar z_1)-uM(1,2)\bar z_1= -uM(1,2) \bar z_1,\nonumber
\\
\frac {dz_2(0)}{dt}&=& +uM(1,2)\bar z_1.
\end{eqnarray}
For $u$ small, this implies that at time $t=1$, 
\begin{equation}\label{AB-simex.3}
z_1(1) \sim \bar z_1(1-uM(1,2)),\quad z_2(1) \sim uM(1,2)\bar z_1.
\end{equation}
Hence,   as long as $z_2(t)$ is small compared to $\bar z_1$,
\begin{equation}\label{AB-simex.4}
\frac {dz_2(t)}{dt}\geq     z_2(t)(r_2-c(2,1)\bar z_1).
 \end{equation}
 and hence exponential growth at rate $(r_2-c(2,1)\bar z_1)\equiv  R>0$ will set in, i.e 
 for $t>1$ and as long as $z_2(t)$ remains small compared to one, 
 \begin{equation}\label{AB-simex.5}
 z_2(t)\sim uM(1,2)\bar z_1 \mathrm{e}^{(t-1)R},
 \end{equation} 
 and so by time $t\sim \frac  1R \ln (uM(1,2)\bar z_1 )$, $z_2$ will have reached a level $O(1)$ 
 that is independent of $u$. Then, for vanishing $u$, the system will evolve over times of order one like 
 the
 mutation free Lotka-Voltera system and approach its unique fixed point 
 $(0,\bar z_2)$. Thus, defining 
 \begin{equation}\label{AB-simex.7}
 Z^u(t)\equiv  (z^u_1(t |\ln u|), z_2^u(t|\ln u|)), 
 \end{equation}
 we see that, in the sense of weak convergence,  
 \begin{equation}
 \label{AB-simex.6}
 \lim_{u\downarrow 0} Z^u(t)= 
 \bar z_1 \1_{0\leq t<1/R}+\bar z_2\1_{t\geq 1/R}.
 \end{equation}
 So, interestingly, on the time scale $\ln (1/u)$, the solution of the deterministic Lotka-Volterra system with 
 mutations converge to a \emph{deterministic jump process}. 
 To my knowledge this scaling was first considered in \cite{AB-BovWang2013} and fully developed in 
 \cite{AB-kraut2018}. A particular situation relating to escape from  an evolutionary stable state was 
 analysed in \cite{AB-BovCoqSma2018}.
 
 What we observed in this simple example is generic and gives rise to the first example of  a \emph{polymorphic evolution sequence} (PES), by which we mean a jump process between 
 equilibria of a sequence of competitive Lotka-Volterra systems. This can be described informally as follows.
 
 Assume (for simplicity) that ${\mathcal X}$ is a countable set. Let $I_0\subset {\mathcal X}$ be a finite set of cardinality $n$ such that
 $LV(n,I_0)$ has an equilibrium $\bar z$ such that $\bar z_i>0$, for all $x_i\in I_0$. 
 
 \noindent\textbf{Step 1:} At time $1$, all the  (mutant) populations at all points $x\not\in I_0$ are either of size 
 zero or of order $u^{\alpha_x}$ with $\alpha_x\in \NN$. The populations at $x\in I_0$ remain close to their equilibrium 
 values. This remains true as long as none of the mutant  populations has reached a level $e>0$
 independent of $u$. 
 
  \noindent\textbf{Step 2:}  The populations of  the types  $x\not \in I_0$ grow exponentially with rate 
  given by their invasion fitness with respect to the resident equilibrium until a time $T_{\epsilon,1}$,
  which is the first time that one of the non-resident populations reaches the value $\epsilon$. 
  Population growth also takes into account mutations. The system is, however, well approximated by a linear system.
  $T_{\epsilon,1}$ is of order $\ln (1/u)$. 
  
    \noindent\textbf{Step 3:} At time $T_{\epsilon,1}$, assume that the set  $J$ of types $y$ for which 
    $\lim_{u\downarrow 0} z_y(T_{\epsilon,1})\neq 0$ is finite (typically, this will be $I_0$ plus one new type). 
    Then, in time of order one, the system will approach the equilibrium of $LV(|J|,J)$. Let $I_1\subset J$ 
    be the subset on which this equilibrium is strictly positive. All types outside $I_1$ have population size
    of some order $u^\alpha$. 
    
     \noindent\textbf{Step 4:} Restart as in Step 2 and iterate.
     
     The general result obtained in \cite{AB-kraut2018} concerns the system of differential equations
     \begin{align}
\frac {d{z}^u_x(t)}{dt}=\left[r(x)-\sum_{y\in\mathbb{H}}\alpha(x,y)z^u_y(t)\right]z^u_x(t)+u\sum_{y\sim x}b(y)m(y,x)z^u_y(t)-u b(x)z^u_x(t)\label{AB-DE},
\end{align}
where $\mathbb {H}$ is the $n$-dimensional hypercube $\{0,1\}^n$, but the same results hold for any 
locally finite graph. The mutations kernel $m(x,y)$ is positive if and only if $x$ and $y$ are connected by an 
edge in $\mathbb{H}$.

     \begin{theorem}[\cite{AB-kraut2018}]\label{AB-MainThm}
Consider the system of differential equations (\ref{AB-DE}).
Assume that the initial conditions $z^u(0)$ are such that for some $\mathbf x^0\subset \mathbb H$, it holds that 
for all $y\in \mathbf x^0$, $z_y^u(0)\sim \bar z_y(\mathbf x^0)$ and for all $y\not\in \mathbf x^0$, $z_y^u(0) \sim u^{\lambda_y}$, 
for some $\lambda_y>0$. Set 
\begin{equation}
\rho^0_y\equiv  \min_{z\in \mathbb H}(\lambda_z+|z-y|), 
\end{equation}
where $|z-y|$ denotes the graph distance between $y$ and $z$,
and 
$T_0\equiv  0$.
 Then  define, for $i\in\mathbb{N}$, 
\begin{align}
y^i_*&\equiv  \arg\hspace{-15pt}\min_{\substack{y\in\mathbb{H}:\\f({y,\mathbf{x}^{i-1}})>0}}\frac{\rho^{i-1}_y}{f({y,\mathbf{x}^{i-1})}},\\
T_i&\equiv T_{i-1}+\min_{\substack{y\in\mathbb{H}:\\ f({y,\mathbf{x}^{i-1})}>0}}\frac{\rho^{i-1}_y}{f({y,\mathbf{x}^{i-1})}},\\
\rho^i_y&\equiv \min_{z\in\mathbb{H}}[\rho^{i-1}_z+|z-y|-(T_i-T_{i-1})f({z,\mathbf{x}^{i-1}})].
\end{align}
Let $\mathbf{x}^i$ be the support of the equilibrium state of the Lotka-Volterra system involving $\mathbf{x}^{i-1}\cup y^i_*$ and set $T_i\equiv \infty$, as soon as there exists no $y\in\mathbb{H}$ such that $f(y,\mathbf{x}^{i-1)}>0$. Then (under some weak non-degeneracy hypotheses), for every $t\notin\{T_i,i\geq0\}$,
\begin{align}
\lim_{u\downarrow 0}z^u({t|\ln u|})=\sum_{i=0}^\infty \1_{T_i\leq t< T_{i+1}}\sum_{x\in\mathbf{x}^i}\delta_{x}\bar{z}_{\mathbf{x}^i}(x).
\end{align}
\end{theorem}

 Let us discuss the meaning of the  terms appearing. First, the quantity is the exponent of $u$ of the population 
 at $y$ at time $1$, i.e. $z_y(1) \sim u^{\rho_y^0}$. This is due to the arrival of mutants from the initial 
 populations at all possible $z$ (noting that in the $u\downarrow 0$ limit, sums are dominate by their maximal 
 terms). The quantity
$ \frac{\rho^{i-1}_y}{f(y,\mathbf{x}^{i-1})}$ is the time (measured in units of $\ln (1/u)$) it takes for the population at $y$ that has initial size $u^{\rho^{i-1}_y}$ that grows with positive  rate $f(y,\mathbf{x}^{i-1})$
to reach size of order $1$. $y^i_*$ is then the type where this happens first, and $T_i$ is the absolute time when this happens.  The formula for the new initial conditions, $\rho^i_y$, is tricky.  It takes into account that
there are three possible sources that could dominate the new initial condition at $y$:
First,  the population at $y$ could have just continued to grow. This gives 
$\rho^i_y=\rho^{i-1}_y +(T_i-T_{i-1})f(y,\mathbf{x}^{i-1})$. 
Second, it could come from mutants form the large populations in $z\in \mathbf{x}^{i-1}$. This gives
$\rho^i_y=0+|z-y|$.  Finally, it could come from the mutants of  any other type $z$, which have grown 
over the last period. This gives
$\rho^i_y=\rho^{i-1}_z+ |z-y|-(T_{i}-T_{i-1})f(z,\mathbf{x}^{i-1})$.

     Theorem \ref{AB-MainThm} is the first instance of a limiting process that describes the
     evolution under the effects of ecology and mutation/migration as  foreseen by adaptive dynamics. 
 The emerging processes      described above can be seen as different cases of 
 \emph{adaptive walks} or  \emph{adaptive flights}. See for instance 
 \cite{AB-KaLe87,AB-Kau92,AB-Orr03, AB-NoKr15,AB-SchKr14,AB-KrugKarl03,AB-JainKrug05,AB-JainKrug07,AB-Jain07,AB-NeidKrug11}.

 \section{The polymorphic substitution sequence}

   A more realistic treatment  requires to take a joint limit when 
     $K\uparrow \infty$ and $u=u_K\downarrow 0$ are taken simultaneously. 

 We first look at this problem under conditions that ensure  that the basic postulate of 
adaptive dynamics, namely that the time scales of \emph{ecology} and 
\emph{evolution}
are well separated, holds. This is the case if the fate of an appearing 
mutant is determined before a new mutant appears. If, as we will
also assume here, the evolutionary advantage of a mutant is positive, 
independent of $K$, the time for a single mutant to produce a number of
offspring of order $K$ will be of order $\ln K$, and the competition with the
resident population will lead close to a new equilibrium in time of order $1$, 
finally, an unfit resident will die out in time of order $\ln K$. 
Thus, to satisfy our assumption, the time between consecutive mutants 
must be larger than $\ln K$, which, given that there are $K$ individuals 
around, means that $u_k\ll (K\ln K)^{-1}$. The results in this chapter are 
based on the paper \cite{AB-CM11} by Champagnat and M\'el\'eard.

\subsection{Heuristics} Under the conditions above, we can expect that the following picture holds for a population that started with an initial condition 
where only a finite number of phenotypes were present.
\begin{itemize}
\item[(i)] For almost all times, the population is  very close to an 
ecological equilibrium where only a finite number $n$ of phenotypes are present. They then determine an (invasion)-fitness landscape. 
\item[(ii)] Mutants that are born from such an equilibrium at a phenotype where
the invasion fitness is negative, die out with probability one. 
\item[(iii)] Mutants that are born from such an equilibrium at a phenotype where
the invasion fitness is positive produce $\epsilon K$ (with $1\gg\epsilon>0$) offspring before they die out 
with strictly positive probability. If they produce this number of offspring, 
this takes time $O(\ln K)$.  
\item[(iv)] From the time when the mutant population has reached the level $\epsilon K$, the population stays close to the solution of the
mutation free deterministic Lotka-Volterra 
system of dimension $n+1$. Under mild hypothesis, this system reaches 
the $\epsilon$-neighbourhood of 
a unique equilibrium with $k\leq n+1$ non-zero components.
\item[(v)] In time of order $\ln K$, the populations corresponding to the 
$n+1-k$ zero-components of this equilibrium die out. 
\end{itemize}

The concept outlined above indicates that the population process that will 
emerge can be seen as a jump process between ecological equilibria of
systems of competitive Lotka-Volterra equations of various dimensions. 
An important and difficult question is what the nature of these equilibria will 
be.

\paragraph{Monomorphism}
We have seen that polymorphic equilibria require specific 
conditions on the coefficients, but  monomorphic equilibria always exist. If we 
start our population process with a monomorphic population at time zero, 
it will thus approach its ecological equilibrium and reach an $\epsilon$-
neighbourhood of it in finite time and will stay there with overwhelming 
probability until a mutant appears. If that mutant appears at a type that has
positive invasion fitness, we have seen above that the population will now 
move towards a unique fixed point of the $2$-dimensional system. 
Now there are two possibilities: either this fixed point is monomorphic, or
it is bi-morphic. If it is monomorphic, it must also be stable, so it cannot be
$(\bar z_1,0)$ (this is unstable by assumption), so it must be $(0, \bar 
z_2)$. This   is stable, if the invasion fitness $f(x_2,x_1)$ is negative. 
Doing the computations, we see that co-existence requires that 
\begin{eqnarray}\label{AB-coex.1}
r(x_2)-\frac{c(x_2,x_1)}{c(x_1,x_1)}r(x_1)&>&0,
\nonumber\\
r(x_1)-\frac{c(x_1,x_2)}{c(x_2,x_2)}r(x_2)&>&0.
\end{eqnarray}
If we assume that the mutants do no differ much from the residents, the 
ratios of the competition kernels in these equations should be very close 
to one. Then, unless $r(x_1)\approx r(x_2)$,  it is not possible that both equations hold simultaneously. 
Thus, the monomorphic fixed point $(0,\bar z_2)$ will be approached and the
population of type $x_1$ will die out. We call this a \emph{trait 
substitution}. This is the generic scenario. The opposite case, when we 
obtain two co-existing types, is called \emph{evolutionary branching}. 
It occurs only if either the two types have almost the same a-priori fitnesses
or it the cross-competition is very weak. 

Starting with a monomorphic initial condition, successive successful 
mutations will thus lead to a sequence of monomorphic populations 
evolving, in some sense, towards higher fitness until
a so-called \emph{evolutionary singularity}. 
The precise convergence of the population process towards such a \emph{trait substitution sequence} was first derived rigorously by Nicolas 
Champagnat \cite{AB-Cha06}.

There are two types of evolutionary singularities that can be met: either,
a trait is reached and a mutation occurs such that coexistence of the 
resident and mutant trait is possible, i.e. evolutionary branching occurs.
The other possibility is that the traits of all possible mutants have 
negative invasion fitness. In that case, the final monomorphic population 
that is reached represents an \emph{evolutionary stable condition} in the 
sense of adaptive dynamics. In that case, evolution appears to come to a 
halt, at least on the time scale of the trait substitution sequence. 
We will see later, in Section \ref{AB-escape}, that this is not the end of the story, and that on longer time-scales, evolution may go on after that. 


We now turn to the rigorous statement concerning the PES, following 
\cite{AB-CM11}).
We begin by defining a (strong) notion of coexisting traits. 

\begin{definition} 
For any $n\geq 2$, we say that the distinct traits  $\mathbf x\equiv (x_1,\ldots ,x_n)$ 
\emph{coexist} if the system $LV(n,  {\mathbf{x}} )$ has a unique non-trivial 
equilibrium $\bar z (\mathbf x)\in (0,\infty)^n$ which is locally strictly stable,  
in the sense  that all eigenvalues of the Jacobian matrix of the system 
$LVS(n,  \mathbf x)$ at $\bar{z} (\mathbf x)$ have strictly negative real 
parts. 
\end{definition}

If the traits $(x_1,\ldots ,x_n)$  coexist, then the \emph{invasion fitness} 
of a mutant trait $ y$ which appears in the resident population is given by the function
\begin{equation}\label{AB-invasion.1} 
f(y,{\mathbf{x}} )\equiv  b(y)-d(y) -\sum_{j=1}^n c(y,x_j)\bar z_j.
\end{equation}
To obtain that the process jumps on the evolutionary time scale from one equilibrium to the next, we need an assumption to prevent cycles, unstable equilibria or chaotic dynamics in the  
deterministic system.

\begin{assumption}\label{AB-conv_to_fixedpoint}
For any given traits $(x_1,\ldots ,x_n)\in {\mathcal X}^n$ 
that coexist and for any mutant trait $y$  
such that $ f(y,{\mathbf{x}} )>0$, there exists a neighbourhood $U$ of  $(\bar z({\mathbf{x}} ),0)$
 such that all solutions of  $LV(d+1,  ({\mathbf{x}} ,y))$ with initial condition in 
 $U\cap (0,\infty)^{n+1}$
converge, as $t \uparrow  \infty$, to a unique locally strictly stable
 equilibrium in $\RR_+^{n+1}$ denoted by $\bar z^{*}((\mathbf x,y))$.
\end{assumption}

We now state the main theorem.

\begin{theorem}\label{AB-PES.0}{\cite{AB-CM11} }Suppose that Assumption \ref{AB-conv_to_fixedpoint} holds. 
Fix $x_1,\dots, x_n$ coexisting traits and assume that the initial conditions 
converge almost surely to $\bar z (\mathbf x)$. 
 Furthermore, assume  that 
\begin{align}\label{AB-Conv_Cond}
\forall V>0, \qquad \exp(-VK)\ll u_K \ll \frac{1}{K\ln(K)}, \qquad \text{as } K\uparrow \infty.
\end{align}
Then, the sequence of the rescaled processes $(\nu^{K}_{ t/ Ku_K})_{t\geq 0}$ with initial state $\nu_0^K$, converges in the sense of finite dimensional distributions to the measure-valued pure jump process 
$\Lambda$ defined as follows:
$\Lambda_0=\sum_{x\in\mathbf x} \bar z_x (\mathbf x) \delta_{x}$
and the process $\:\Lambda\:$ jumps for all $y\in \mathbf x$  from
 \begin{equation}
 \sum_{x\in\mathbf x} \bar{z}_{x}(\mathbf x) \delta_{x}
\quad \text{ to } \quad
  \sum_{x\in \mathbf x\cup (x+h)} \bar z^*_{x}(\mathbf x\cup  y) \delta_{x}
  \end{equation}
  with infinitesimal rate
  \begin{equation}
  \sum_{x\in \mathbf x} m(x)b(x) \bar{z}_{x} (\mathbf x)    \frac{ f(y,\mathbf x)_+}{b(y)}M(x, dy).
  \end{equation}
  The process $\L$ is called the \emph{polymorphic evolution sequence (PES)}.
\end{theorem}

\begin{remark} In Reference \cite{AB-CM11}, the mutation kernel is assumed absolutely continuous, but this 
assumption is not necessary, as one can easily check.
\end{remark}

 The limiting process described in the theorem is called the \emph{polymorphic evolution sequence (PES)}.
 A special case is the \emph{trait substitution sequence (TSS)}, when all equilibria are monomorphic. 
 In some sense  the situation that the dimension of the successive equilibria stays constant is generic.
 Cases when the dimension increases are called \emph{evolutionary branching}.

\subsection{The canonical equation} 

The trait substitution sequence still contains $\sigma$, the scale of a mutation step, as a small parameter. 
If we denote the corresponding process by $X^\sigma$, one can obtain a further limiting process which describes 
continuous evolution of the population in phenotypic space. In adaptive dynamics, this equation is called the 
\emph{canonical equation}.

\begin{theorem}[Remark 4.2 in \cite{AB-CM11}]
\label{AB-1.2.2thm} 
If Assumption \ref{AB-ass} is satisfied and the family of initial states of the rescaled TSS, $X^{\sigma}_0$, is bounded in $L^2$ 
and converges to a random variable $X_0$, as $\sigma \rightarrow 0$,
then, for each $T>0$, the rescaled TSS $X^{\sigma}_{t/\sigma^{2}}$ converges, as $\sigma\downarrow 0$, 
in the Skorokhod topology on $\mathbb D([0,T],\mathcal X)$ 
 to the process $(x_t)_{t\leq T}$ with initial state $X_0$ and with deterministic sample path, which is the
 unique solution of an ordinary differential equation, known as CEAD:
\begin{equation}\label{AB-(CEAD)}
\frac {d x_t}{dt}=\int h\:[h\:m(x_t)\:\overline z(x_t)\:\partial_1 f(x_t,x_t)]_+ M(x_t,dh) ,
\end{equation} 
where $\partial_1f$ denotes the partial derivative of the function $f(x,y)$ with respect to the first variable $x$.
\end{theorem} 

\begin{remark} In Reference \cite{AB-CM11}, the mutation kernel is assumed absolutely continuous, but this 
assumption is not necessary, as one can easily check.
\end{remark}
Note that the CEAD has fixpoints where the derivative of $f(x,x)$ vanishes. Typically, a population will evolve 
towards such a fixpoint and slow down. The further fate of the population cannot be determined on the basis of 
the CEAD alone. However, in the underlying stochastic model, the population can either stay fixed, if the fixpoint is stable and an evolutionary stable situation is reached, or, in case of an unstable fixpoint, 
evolutionary branching may occur.

\section[In one step]{To the CEAD in one step}

Deriving at the CEAD through the successive limits $K\uparrow\infty, u_K\downarrow 0$ first, and $\sigma\downarrow 0$ later is somewhat unsatisfactory. It would be more natural to give conditions under which 
the limits of  large population size, $K\rightarrow\infty$, rare mutations, 
$u_K \rightarrow 0$, 
and small mutation steps, $\sigma_K\rightarrow 0$, can be 
taken \emph{simultaneously} and lead to the CEAD. 
Such a result was achieved in a paper with  M. Baar and N. Champagnat \cite{AB-B14}.
It turns out that the combination of the three limits simultaneously entails some considerable technical 
difficulties. The fact that the mutants have only a $K$-dependent small evolutionary advantage  
decelerates the dynamics of the microscopic process such that the time of any 
macroscopic change between resident and mutant diverges with $K$. 
This makes it impossible to use a law of large numbers 
 to approximate the stochastic system with the corresponding deterministic system during the time of invasion.
Showing  that the stochastic system
 still follows in an appropriate sense the  
 corresponding competition Lotka-Volterra system (with $K$-dependent coefficients)
 requires a completely  new approach. Developing this approach, which can be seen 
 as a rigorous "stochastic Euler-scheme", is the main novelty in the paper  \cite{AB-B14}. The proof requires methods, based on 
 couplings 
 with discrete time Markov chains combined with some standard potential theory arguments for the "exit from a domain problem" 
in a moderate deviations regime, as well 
 as comparison and convergence results of branching processes.

\subsection{The main result} \label{AB-result}
In this section, we present the main result of  \cite{AB-B14}, namely the convergence to the canonical equation of adaptive dynamics in one step. 
The time scale on which we control the population process
 is $t/(\sigma_K^2u_K K)$. 
For technical reasons, we make the following simplifying assumption.
\begin{assumption}\label{AB-ass3}
\begin{itemize}
\item[(i)] The trait space $\cX$ is a  subset of $\RR$.
\item[(ii)] The mutant distribution $M(x,dh)$ is  atomic and the number of atoms is uniformly bounded.
\item[(iii)] For all $x\in\mathcal X$, \quad$\partial_1 f(x,x)\neq 0$.
\end{itemize}
\end{assumption}
Assumption \ref{AB-ass3} implies that either  $\forall x\in\mathcal X$: $\partial_1 f(x,x)>0$ or  $\forall x\in\mathcal X$:$\partial_1 f(x,x)<0$. Therefore coexistence of two traits is not possible. Without loss of generality we can assume that, $\forall x\in\mathcal X$, \:$\partial_1 f(x,x)>0$. In fact, a weaker assumption is sufficient, see Remark \ref{AB-remark_main_thm}.(iii).  
\begin{theorem} \label{AB-main_thm}
Assume that Assumptions \ref{AB-ass} and \ref{AB-ass3} hold and that there exists a small $\alpha>0$ such that
\begin{align}
\label{AB-conv1}  		K^{- \frac  1 2 +\alpha}&\ll \sigma_K\ll 1 \qquad \qquad \text{ and }  \\
\label{AB-conv2}	\qquad\qquad	\exp(-K^{\alpha})&\ll u_K\ll \frac{\sigma_K^{1+\alpha}}{K\ln K},	\quad \text{ as } \qquad K\rightarrow \infty.
\end{align} 
Fix $x_0\in\mathcal X$ and let $(N^{K}_0)_{K\geq 0}$ be a sequence of $\mathbb N$-valued random variables such that
$N^{K}_0 K^{-1}$ converges in law, as $K\to\infty$,  to the positive constant $\overline z(x_0)$ and
is bounded in $\mathbb L{}^p$, for some $p > 1$. \\[0.5em]
For each $K\geq 0$,
let $\nu^{K}_{t}$ be the  process generated by $\mathcal L^K$ with monomorphic initial state ${N^{K}_0} K^{-1} \delta_{\{x_0\}}$.
Then, for all $T>0$, the sequence  of rescaled processes, $\big(\nu^{K}_{t / (Ku_K \sigma_K{}^2)}\big)_{0\leq t\leq T}$,
converges in probability, as $K\rightarrow \infty$,  with respect to the Skorokhod topology 
on  $\mathbb D([0,T],\mathcal M(\mathcal X))$ to the measure-valued process $\overline z(x_t)\delta_{x_t}$, 
where $(x_t)_{0\leq t\leq T}$ is given as a solution of the CEAD, 
\begin{equation}\label{AB-CEAD}
\frac {d x_t}{dt}=\int_{ \mathbb Z} h\:[h\:m(x_t)\:\overline z(x_t)\:\partial_1 f(x_t,x_t)]_+ M(x_t,dh),
\end{equation}
with initial condition $x_0$.  
\end{theorem}
\begin{remark}\label{AB-remark_main_thm}
\begin{itemize}
\item[(i)] If $x_t\in\partial \mathcal X$ for $t>0$, then (\ref{AB-CEAD}) is $\frac {d\:x_t}{dt}=0$, i.e.,  the process stops.
\item[(ii)] The condition $ u_K\ll \frac{\sigma_K^{1+\alpha}}{K \ln K}$ allows mutation events during an invasion phase of a mutant trait,
	see below, but ensures that there is no \emph{successful} mutational event during this phase.
\item[(iii)] The fluctuations of the resident population are of order $K^{- \frac  1 2}$, 
	so the condition $K^{- \frac  1 2 +\alpha}\ll \sigma_K$ ensures that the sign
	of the initial growth rate is not influenced by these fluctuations.
\item[(iv)]  $\exp(K^{\alpha})$ is the time the resident population stays with high probability in an $O(\epsilon\sigma_K)$-neighbourhood of an attractive domain. 
	This is a moderate deviation result.
	Thus the condition $\exp(-K^{\alpha})\ll u_K$ 
	ensures that the resident population is still in this neighbourhood when a mutant occurs.
\item[(v)] The time scale is  $(Ku_K \sigma_K{}^2)^{-1}$ since the expected time for a mutation event is $(K u_K)^{-1}$, 
	the probability that a mutant invades
	is of order $\sigma_K$ and one needs $O(\sigma_K^{-1})$ mutant invasions to see an $O(1)$ change of the resident trait value. 
	 
\end{itemize}
\end{remark}

\subsection{The structure of the proof of Theorem \ref{AB-main_thm}}\label{AB-The main idea}

Under the conditions of the theorem, the evolution of the population will be described as 
a succession of \emph{mutant invasions}.

\noindent\textbf{Analysis of a  single invasion step:} 
In order to analyse the invasion of a mutant, we divide the time until a mutant trait has fixated in the population into two phases.

\subparagraph{Phase 1} Fix  $\epsilon>0$ and prove the existence of a constant, $M<\infty$, independent of $\epsilon$, such that, as long as all mutant densities are smaller than $\epsilon\sigma_K$, 
the resident density stays in an $M\epsilon\sigma_K$-neighbourhood of  $\overline z(x)$. Note that, 
because  mutations are rare and the population size is large, 
the monomorphic initial population has  time to stabilise in an $M\epsilon\sigma_K$-neighbourhood of this equilibrium $\overline z(x)$
before the first mutation occurs.
(The time of stabilisation is of order $\ln(K)\sigma_K^{-1}$ and the time where the first mutant occurs is of order $1/Ku_K$).

This allows  to approximate the number of the mutants of trait $y_1$ by a branching process with birth 
rate $b(y_1)$ and death rate $d(y_1)-c(y_1,x)\overline z(x)$ such that we can 
compute the probability that the number of the mutant  reaches $\epsilon 
\sigma_K$, which is of order $\sigma_K$, as well as the time it takes to reach this level or 
to die out. Therefore, the process  needs $O(\sigma_K^{-1})$ mutation events 
until there appears a mutant subpopulation which reaches a size
$\epsilon \sigma_K$. Such a mutant is called \emph{successful mutant} and its trait will 
be the next resident trait. 
  
\subparagraph{Phase 2} If a mutant population with 
 trait $y_s$  reaches the size 
$\epsilon \sigma_K$, it will increase to an  $M\epsilon\sigma_K$-neighbourhood of its equilibrium density
 $\overline z(y_s)$. 
Simultaneously, the density of the resident trait decreases to $\epsilon \sigma_K$ and finally dies out. %
Since the fitness advantage of the mutant trait is only of order $\sigma_K$, the dynamics 
of the population process and the corresponding deterministic system 
are very slow, and require  a time of order at least $\sigma_K^{-1}$ to reach an 
$\epsilon$-neighbourhood of its  
equilibrium density.   
Thus, the 
 law of large numbers, see Theorem \ref{AB-lil.1} cannot be used to control this phase, as it covers only finite, $K$, independent time intervals.
The method we develop to handle this situation can be seen as a rigorous 
 stochastic "Euler-Scheme".
Nevertheless, the proof contains an idea which is strongly connected with the properties of the deterministic 
dynamical system. Namely, the deterministic system of equations for the case $\sigma_K=0$ has 
an invariant manifold of fixed points  with a vector field independent of $\sigma_K$ pointing towards this manifold. Turning on 
a small 
$\sigma_K$, we therefore expect the stochastic system to stay close to this invariant manifold and to move 
along it with speed of order $\sigma_K$. 
With this method one can prove that the mutant density reaches the $M\epsilon\sigma_K$-neighbourhood of $\overline z(y_s)$ and
 the resident trait dies out.

\noindent\textbf{Convergence to the CEAD:} 
The proof of convergence to the CEAD uses comparison of the measure valued process $\nu^K_t$ with two families 
of control 
processes,  $\mu^{1,K,\epsilon}$ and $\mu^{2,K,\epsilon}$, which converge to the CEAD as $K\rightarrow\infty$ and then $\epsilon\rightarrow 0$. To make more precise statements, one uses the  order relation $\preccurlyeq $ for random variables. Roughly speaking, $X\preccurlyeq Y$ will mean that $Y$ is larger than $X$ in law.

Given $T>0$, with the results of the two invasion phases, 
 one  defines, for all $\epsilon>0$,  two measure-valued processes, 
 in $\mathbb D([0,\infty),\mathcal M(\mathcal X))$,  such that, for all $\epsilon>0$,
\begin{equation}
\lim_{K\rightarrow \infty} \mathbb P\left [\forall\: t\leq \tfrac T{Ku_K\sigma_K^2}:\quad \mu_t^{1,K,\epsilon} \preccurlyeq \nu_t^{K} \preccurlyeq \mu_t^{2,K,\epsilon} \:\right] =1,
\end{equation}
 and, for all $\epsilon>0$ and $i\in\{1,2\}$,
\begin{equation}
\lim_{K\rightarrow \infty}
		 \mathbb P\left [\:\sup_{0\leq t\leq T/(Ku_K\sigma_K^2)} \Big\Vert \: \mu^{i,K,\epsilon}_{t/(K u_K \sigma_K{}^2)}-
		 \overline z(x_t)\delta_{x^{}_t}\:\Big\Vert^{}_0>\delta(\epsilon) \:\right] = 0,
\end{equation}
for some function $\delta$ such that $\delta(\epsilon)\rightarrow 0$ when $\epsilon\rightarrow 0$.

Here $\|\cdot\|_0$ denotes the Kantorovich-Rubinstein norm:
\begin{equation}
	\Vert \,\mu_t\, \Vert^{}_0\equiv\sup\left\{\int_{\mathcal X} f d\mu_t: f\in \text{Lip}_1(\mathcal X) \text{ with } \sup_{x\in\mathcal X}|f(x)|\leq 1\right\},
\end{equation}		
	where $\text{Lip}_1(\mathcal X)$  is the space of Lipschitz continuous functions from $\mathcal X$ to $\mathbb R$ 
	with Lipschitz norm one (cf. \cite{AB-B_MT} p. 191). 	
This implies   the theorem.

\section{Escape through a fitness well}\label{AB-escape}

\newcommand\Ninterval[2]{[[ #1,#2]]}

\newcommand{\ind}[1]{{\mathbf{1}}_{\left[ {#1} \right] }}

When a population reaches an ESC, there may still be uninhabited loci where the invasion fitness is positive
but that cannot be reached by a single mutation from the current population. 
This question was already addressed by Maynard-Smith \cite{AB-Maynard70} and heuristic computations of 
the crossing time of such fitness valleys were computed by Gillespie \cite{AB-gillespie1984molecular}.
In \cite{AB-BovCoqSma2018} we 
have analysed how  such a fitness valley can be crossed in a simple scenario where the
trait space is the finite set $\{0,\dots, L\}$,  the  resident 
population is monomorphic with trait zero, the invasion fitness is negative for $1,\dots. L-1$ and positive for 
$L$. 

In contrast to the previous chapters, we analyse a wider range of 
dependencies of the mutation rate on the carrying capacity, interpolating 
all the extreme regimes $K\uparrow \infty $ first, $u\downarrow 0$ later, 
to the regime $u\ll 1/(K\ln K)$. 
As we will show, three essentially different regimes  occur. In the first, 
 the mutation rate is so  
large that  many mutants (a number of order $K$) are created in 
a time of order $1$. In this case the fixation time scale is dominated by 
the time needed for a successful mutant to invade (which is of order 
$\log K$).
The second scenario occurs if the mutation rate is smaller,  but large 
enough so that a fit mutant will appear before the resident population dies 
out. In this case the fixation time scale is
exponentially distributed and
dominated by the time needed 
for the first successful mutant to be born. The last possible scenario is the
 extinction of the population before the fixation of the fit mutant, 
which occurs when the mutation rate is very small 
(smaller than $\mathrm{e}^{-CK}$ for a constant $C$ to be made precise later).

\subsection{The setting}

We analyse the escape problem in a specific simple special case of the
general model, that does,  however,
capture the key mechanism.
We choose the trait  space  $\mathcal{X}\equiv   \{0, 1,\ldots,L\}$. For each trait $i$ we denote by $X_i(t)$ the number of individuals of trait $i$ at time $t$.

For simplicity, we allow mutations only in the forward directions and to nearest neighbours, that is, we set
$m_{ij}= u\delta_{i+1,j}$.

For $n,m\in\NN_0$ such that $n\leq m$, we introduce the notation $[[ n,m]]\equiv  \{n,n+1,\ldots,m\}$. 
We want to consider the situation when an equilibrium population at $0$
is an evolutionary stable condition, and when $L$ is the closest trait with 
a positive invasion fitness:
\begin{assumption}\label{AB-ass.1}\emph{}\\
\begin{enumerate}
	\item[$\bullet$] (Fitness valley) All traits are unfit with respect to 0 except $L$:
	\begin{equation} \label{AB-A1} 
		f(i,0)<0 \text{ for } i\in[[ 1,L-1]]  \text{ and } f(L,0)>0.
	\end{equation}
	\item[$\bullet$] All traits are unfit with respect to $L$:
	\begin{equation} \label{AB-A2}
		f(i,L)<0 \text{ for  } i\in[[ 0,L-1]].
	\end{equation}
	\item[$\bullet$]  The following fitnesses are different:
	\begin{align}
		f(i,0)&\neq f(j,0) \text{ for all }i\neq j,\\
		f(i,L)&\neq f(j,L) \text{ for all }i\neq j.
	\end{align}
\end{enumerate}
\end{assumption}

Under these assumptions, all mutants created by the initial population 
initially have a negative growth rate  and thus tend to die out. However, 
if  such mutants survive long enough to give rise to further mutants, etc, such that 
eventually
an individual will reach the trait $L$, it will found a population at this trait that, with 
positive probability,
will grow and possibly eliminate the resident population through competition. 

\subsection{Results} 
The more interesting results in \cite{AB-BovCoqSma2018} concern the case when 
$ u= u_K\sim K^{-1/\alpha}$. There are two very different cases to distinguish.
First, if $K u^L=K^{1-L/\alpha}\gg  1$ (i.e. $\alpha>L$), there will be essentially immediately a divergent number of mutants at  $L$. These the grow exponentially with rate $f(L,0)$ and therefore will reach a macroscopic 
level at time of order $\ln K/(\alpha f(L,0))$.  

Second, if $K u^L=K^{1-L/\alpha}\ll 1 $ (i.e. $\alpha<L$), there are typically no mutants at $L$. 

For $v\geq 0$ and $0\leq i \leq L$,   let  $T_v^{(K,i)}$ denote  the first  time the $i$-population  reaches the  size 
$\lfloor v K\rfloor$, 
\begin{equation}\label{AB-defTepsKM}
 T^{(K,i)}_v \equiv  \inf \{ t \geq 0, X_i(t)= \lfloor v K \rfloor \}.
\end{equation}

Let us introduce 
\begin{equation}\label{AB-timetoextinction}
 t(L,\alpha)\equiv  \frac{L}{\alpha} \frac{1}{f(L,0)}+ \sup \left\{\left(1 - \frac{i}{\alpha}\right) \frac{1}{|f(i,L)|}, 0 \leq i \leq L-1\right\}, 
 \end{equation}
and the time needed for the   populations of all types but $L$ to get extinct,
\begin{equation}\label{AB-defT0ttsaufM}
 T^{(K,\Sigma)}_0 \equiv  \inf \Big\{ t \geq 0, \sum_{0\leq i \leq L-1}X_i(t)= 0 \Big\}.
\end{equation}
With this notation we have the following asymptotic result.

\begin{theorem}
 \label{AB-pro_phase1_mugrand}
Assume that $L< \alpha < \infty$.
 Then there exist two positive constants $\varepsilon_0$ and $c$ such that, for every $0<\varepsilon \leq \varepsilon_0$,
 \begin{equation}\label{AB-eq1th}
  \liminf_{K \to \infty} \PP \left( (1-c\varepsilon)\frac{1}{\alpha} \frac{L}{f(L,)}< 
  \frac{T^{(K,L)}_\varepsilon }{\log K} { <\frac{T^{(K,L)}_{\bar{x}_L-\varepsilon} }{\log K}}
 < (1+c\varepsilon)\frac{1}{\alpha} \frac{L}{f(L,0)} \right) \geq 1-c\varepsilon.
 \end{equation}
{Moreover,
 \begin{equation}\label{AB-eq2th}
    \frac{T^{(K,\Sigma)}_{0}}{\log K} \to t(L,\alpha), \quad \text{in probability}, \quad (K \to \infty) 
 \end{equation}
and} there exists a positive constant $V$ such that
\begin{equation} \label{AB-eq3th}   
\limsup_{K \to \infty} \PP \left( \sup_{t \leq \mathrm{e}^{KV}}\left| 
 X_L \left(T^{(K,L)}_{\bar{x}_L-\varepsilon}+t \right)-\bar{x}_L K  \right|
 > c\varepsilon K\right) \leq c\varepsilon. \end{equation}
 \end{theorem}

In other words, it takes a time of order $t(L,\alpha) \log K$ for the $L$-population to outcompete the other populations and enter in a neighbourhood 
of its monomorphic equilibrium size $\bar{x}_L K$. Once this has happened,  it stays close to this equilibrium for at least a time $\mathrm{e}^{KV}$, where 
$V$ is a positive constant.

Note that the constant $t(L,\alpha)$ can be intuitively computed from the deterministic limit. Indeed, 
for $\alpha>L$, we will prove that the system performs small 
fluctuations around the deterministic evolution studied above: the $i$-population first stabilises around {$O(K u^i)$} in a time of order one, then 
the $L$-population grows exponentially with rate $f(L,0)$ and needs a time of order $L\log K /(\alpha f(L,0))$ to reach 
a size of order $K$,  while the other types stay stable, 
the swap between populations $0$ and $L$ then {takes} a time of order one, and finally, for $i\neq L$, the $i$-population decays exponentially 
from {$O(K u^i)$} 
to extinction with a rate given by the lowest (negative) fitness of its 
left neighbours ({sub-critical branching process,} needs a time close to $(\sup_{j\in\Ninterval{0}{i}}(1-j/\alpha)/|f(i,L)|)\log K$). Thus the time until 
extinction of all non-$L$ populations is close to a constant  times $\log K$.\\

Next we consider the case  when $L/\alpha>1$. In this case, there is no $L$-mutant
at time one, and the fixation of the trait $L$ happens on a much longer time scale. In fact, there 
will be some last  $j<L$ where there will be of order $\gg 1$ mutants present essentially 
all the time. Already at  $j+1$, mutants arrive only sporadically and will typically get extinct quickly. 
Mutants arrive at  $L$ only when the rare event that a sequence of mutants manages to survive 
the trip from $j$ to $L$ occurs. Such an excursion can be described as follows:
First, a mutant of type $j+1$ is born from the $j$-population. This generates 
a subcritical branching process with  birthrate $b_{j+1}$ and death rate $d_{j+1}+c_{j+1,0} \bar x_0$. Define the parameter $\rho_{j+1}\equiv 
b_{j+1} /(b_{j+1}+d_{j+1}+c_{j+1,0}\bar x_0)$. 
The expected number of individuals that are generated by this process before extinction is 
then
\begin{equation}
\lambda(\rho_{j+1})=\sum_{k+1}^\infty \frac {(2k)!}{(k-1)!(k+1)!}\rho_{j+1}^k(1-\rho_{j+1})^{k+1}.
\end{equation}
Thus, on average, the probability that during the lifetime of the descendants of this mutant 
a $j+2$ mutant is born is $ u \lambda(\rho_{j+1})$. Should that happen, 
this will create a subcritical process of $j+2$ individuals, that produce a 
$j+3$-mutant with average probability $ u\lambda(\rho_{j+2})$, and so forth. 
Thus we see that the probability that a $j+1$-child of the $j$-population has offspring that reaches 
$L$ is about $ u^{L-j} \prod_{i=j+1}^{L-1} \lambda(\phi_i)$. This explains the result stated in the theorem below.

\begin{theorem} \label{AB-pro_mupetit}
\begin{itemize}
 \item[$\bullet$] {
 Assume that  $\alpha \notin \NN$ and $\alpha < L$.}
 Then there exist {two} positive constants $\varepsilon_0$ and $c$, and two exponential random variables
  $E_\mp$  
 with  parameters   
 \begin{equation}
  (1\pm c\varepsilon){\frac{ \bar{x}_0b_0... b_{\lfloor \alpha \rfloor-1}}{|f(1,0)|...|f({\lfloor \alpha \rfloor, 0})|}} \frac{f(L,0)}{b_L} \prod_{i=\lfloor \alpha \rfloor +1}^{L-1}\lambda(\rho_i)
\end{equation}
 such that, for every $\varepsilon \leq \varepsilon_0$,
 \begin{equation}
  \liminf_{K \to \infty} \PP \left( E_-\leq 
  {T^{(K,L)}_{\bar{x}_L-\varepsilon} \vee T^{(K,\Sigma)}_{0}}{K u^L} \leq 
  E_+ \right) \geq 1-c\varepsilon.
 \end{equation}
 \item[$\bullet$] {There exists a positive constant $V$ such that if $ u$ satisfies 
 \begin{equation}
  \mathrm{e}^{-VK} \ll K u \ll  1 ,
  \end{equation}
 then the same conclusion holds, with the corresponding parameters, for $E_-$ and $E_+$:
 \begin{equation}
  (1+c\varepsilon)\bar{x}_0 \frac{f(L,0)}{b_L} \prod_{i=1}^{L-1}\lambda(\rho_i)
\qquad \text{and} \qquad 
(1-c\varepsilon)\bar{x}_0 \frac{f(L,0)}{b_L} \prod_{i=1}^{L-1}\lambda(\rho_i).
\end{equation} }
\end{itemize}
Moreover, {under both assumptions,} there exists a positive constant $V$ such that
\begin{equation}   
\limsup_{K \to \infty} \PP \left( \sup_{t \leq \mathrm{e}^{KV}}\left| 
 X_L \left(T^{(K,L)}_{\bar{x}_L-\varepsilon}+t \right)-\bar{x}_L K  \right|
 > c\varepsilon K\right) \leq c\varepsilon. 
 \end{equation}
 \end{theorem}

In the {first case,} the typical trajectories of the process are as follows: mutant populations of 
type $i$, for $1 \leq i \leq \lfloor \alpha \rfloor$, 
reach a size of order $K u^i \gg 1$ in a time of order $\log K $ (they are well approximated by birth-death processes with immigration and their behaviour is then close to the deterministic limit), and mutant populations of type $i$, 
for $\lfloor \alpha \rfloor 
+1 \leq i \leq L$, describe a.s.\ finite excursions, 
among which a proportion of order $ u$ produces a mutant of type $i+1$. 
Finally, every $L$-mutant has a probability $f(L,0)/b_L$ to produce a population which 
outcompetes all 
 other populations.
 The term $\lambda(\rho_i)$ is the expected number of individuals in an excursion of a subcritical
  birth and death process of birthrate $b_i$ and death rate $d_i+c_{i0}\bar x_0$ {excepting the first individual.}
  Hence $ u \lambda(\rho_i)$ is the approximated probability for a type $i$-population 
  $(\lfloor \alpha \rfloor +1 \leq i \leq L-1)$ to produce a mutant of type $i+1$,
{and} the overall time scale can be recovered as follows:
  \begin{enumerate}{
\item The last 'large' population is the $\lfloor \alpha \rfloor$-population, which {reaches} a size of order 
{$K u^{\lfloor \alpha \rfloor}$} after a time
   which does not go to infinity with $K$.
   \item The $\lfloor \alpha \rfloor $-population produces an excursion of an  $(\lfloor \alpha \rfloor+1 )$-population at a rate of order 
{$K u^{\lfloor \alpha \rfloor+1}$, which} has a probability of order $ u$ to produce an excursion of a  
$(\lfloor \alpha \rfloor+2 )$-population, and so {on,}}
\end{enumerate}
giving the order {$K u^L$}.\\

{Notice that Theorem \ref{AB-pro_mupetit} implies that, for any mutation rate which converges to zero more slowly than 
$\mathrm{e}^{-VK}/K$, the population will cross 
the fitness valley with a probability tending to 1 as $K\to\infty$. Our results thus cover a wide range of biologically relevant 
cases.}

%
%
%

\medskip

\noindent\textbf{Acknowledgements.} I am indebted to all my collaborators on this 
project, notably Martina Baar, Nicolas Champagnat, Loren Coquille, Anna Kraut, Rebecca Neukirch, and
Charline Smadi.   Many thanks for insightful discussions on adaptive walks with Joachim Krug. I am particularly grateful to our colleagues from medicine, 
Nicole Glodde, Michael Hölzel, Meri Rogova, and Thomas Tüting for fruitful and inspiring collaborations.

%

\end{document}